\documentclass[3p,times,procedia]{elsarticle}
\usepackage{nupha_ecrc}

\volume{00}

\firstpage{1}

\journalname{Nuclear Physics A}

\runauth{Kari J.\ Eskola, Ilkka Helenius, Petja Paakkinen and Hannu Paukkunen}

\jid{nupha}

\jnltitlelogo{Nuclear Physics A}

\usepackage{amssymb}
\usepackage[figuresright]{rotating}

\usepackage[T1]{fontenc}
\usepackage[utf8]{inputenc}
\usepackage[english]{babel}
\usepackage{amsmath}
\usepackage{floatflt}
\usepackage{microtype}

\begin{document}

\begin{frontmatter}



%% Instructions from Editor: Please use the following \dochead only in the preprint version (e-print arXiv etc.);
%% use empty \dochead{} when submitting to Nuclear Physics A!
\dochead{XXVIIIth International Conference on Ultrarelativistic Nucleus-Nucleus Collisions\\ (Quark Matter 2019)}
%% e.g. \dochead{17th International Conference on Dynamical Processes in Excited States of Solids}

\title{Impact of dijet and D-meson data from 5.02 TeV p+Pb collisions on nuclear PDFs}


\author[jyu,hip]{Kari J.\ Eskola}
\author[jyu,hip]{Ilkka Helenius}
\author[igfae,fn1]{Petja Paakkinen}
\author[jyu,hip]{Hannu Paukkunen}

\address[jyu]{University of Jyvaskyla, Department of Physics, P.O. Box 35, FI-40014 University of Jyvaskyla, Finland}
\address[hip]{Helsinki Institute of Physics, P.O. Box 64, FI-00014 University of Helsinki, Finland}
\address[igfae]{Instituto Galego de F{\'\i}sica de Altas Enerx{\'\i}as IGFAE, Universidade de Santiago de Compostela, E-15782 Galicia-Spain}
\fntext[fn1]{Speaker.}

\begin{abstract}
We discuss the new constraints on gluon parton distribution function (PDF) in lead nucleus, derivable with the Hessian PDF reweighting method from the 5.02 TeV p+Pb measurements of dijet (CMS) and $D^0$-meson (LHCb) nuclear modification ratios. The impact is found to be significant, placing stringent constraints in the mid- and previously unconstrained small-$x$ regions. The CMS dijet data confirm the existence of gluon anti-shadowing and the onset of small-$x$ shadowing, as well as reduce the gluon PDF uncertainties in the larger-$x$ region. The gluon constraints from the LHCb $D^0$ data, reaching down to $x \sim 10^{-5}$ and derived in a NLO perturbative QCD approach, provide a remarkable reduction of the small-$x$ uncertainties with a strong direct evidence of gluon shadowing. Furthermore, we find a good description of the data even down to zero $D^0$-meson transverse momentum within a purely DGLAP-based approach without a need for imposing any non-linear effects. Importantly, the constraints obtained from the dijet and $D^0$ data are mutually fully consistent, supporting the universality of nuclear PDFs in hard-scattering processes.
\end{abstract}


\end{frontmatter}

\section{Introduction}
\label{sec:intro}

The gluon content of nuclei, relevant for many observables in nucleus--nucleus collisions, has for a long time been one of the least well known aspects of the nuclear parton distribution functions (nPDFs). This has stemmed from the persistent lack of experimental data which would reliably probe the nuclear gluon densities. Prior to the advent of the LHC, the only such data available were the inclusive pion-production measurements in deuteron--gold collisions at RHIC, which provided evidence of gluon antishadowing (enhancement in the bound-nucleon PDF at the momentum fraction around $x \sim 0.1$ with respect to that of the free proton), but did not have a sufficient $x$ reach to probe the possible gluon shadowing (the respective suppression at smaller $x$). In EPPS16~\cite{Eskola:2016oht}, the first global nPDF analysis to include LHC data, additional constraints were obtained from CMS measurements of dijet production in 5.02 TeV proton--lead (p+Pb) collisions. At the time of the analysis no proton--proton (p+p) baseline at the same collision energy was available, however. To reduce the dependency on free-proton PDFs, the study had to resort to taking the forward-to-backward ratio of the measured cross sections, loosing some information in the process. The later 5.02 TeV p+p data taking has enabled experiments to measure nuclear modification ratios of hadronic observables, providing more direct probes for the nuclear modification of the gluon PDFs. Here, based on Refs.~\cite{Eskola:2019dui,Eskola:2019bgf}, we discuss the impact of the the CMS dijet~\cite{Sirunyan:2018qel} and LHCb $D^0$~\cite{Aaij:2017gcy} measurements on the nPDFs, assessed in terms of Hessian PDF reweighting~\cite{Paukkunen:2014zia}. In particular, the $D^0$ production is calculated within the SACOT-$m_{\rm T}$~\cite{Helenius:2018uul} general-mass variable flavour number scheme (GM-VFNS), providing a realistic estimate of the probed $x$ region.

\vspace{-0.25cm}
\section{Constraints from CMS self-normalized dijet production data}
\label{sec:dijet}

\begin{figure}
  \includegraphics[width=\textwidth]{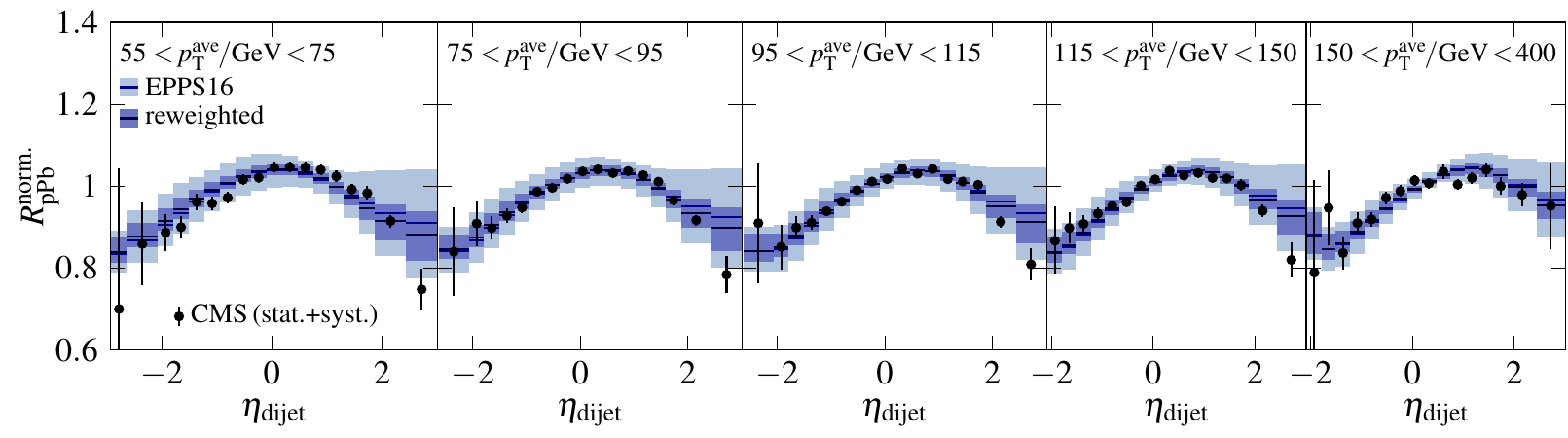}
  \vspace{-0.6cm}
  \caption{Comparison of the CMS self-normalized dijet nuclear modification ratio with NLO calculations using the EPPS16 nPDFs before and after reweighting with the data. Figure from Ref.~\cite{Eskola:2019dui}.}
  \label{fig:RpPb_dijet}
\end{figure}

\begin{floatingfigure}[r]{0.41\textwidth}
  \centering
  \includegraphics[width=0.41\textwidth]{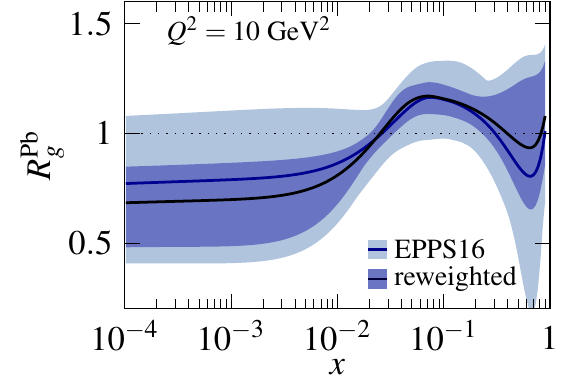}
  \vspace{-0.75cm}
  \caption{The impact of reweighting with the CMS self-normalized dijet nuclear modification ratio on EPPS16 gluon PDF modification. Figure from Ref.~\cite{Eskola:2019dui}.}
  \label{fig:EPPS16_rw_dijet}
\end{floatingfigure}

The CMS data of dijet nuclear modification ratio~\cite{Sirunyan:2018qel} is provided in terms of a double ratio
\begin{equation}
  R_{\rm pPb}^{\rm norm.} = \frac{\frac{1}{\mathrm{d}\sigma^{\rm p+Pb}/\mathrm{d}p_\mathrm{T}^{\rm ave}}\,\mathrm{d}^2\sigma^{\rm p+Pb}/\mathrm{d}p_\mathrm{T}^{\rm ave}\mathrm{d}\eta_{\rm dijet}}{\frac{1}{\mathrm{d}\sigma^{\rm p+p}/\mathrm{d}p_\mathrm{T}^{\rm ave}}\,\mathrm{d}^2\sigma^{\rm p+p}/\mathrm{d}p_\mathrm{T}^{\rm ave}\mathrm{d}\eta_{\rm dijet}},
\end{equation}
where the self-normalization of the p+Pb and p+p cross sections is introduced to cancel part of the experimental correlated uncertainties. We have verified that also the free-proton PDF and scale uncertainties efficiently cancel in this observable~\cite{Eskola:2019dui}. Here, $p_\mathrm{T}^{\rm ave}$ and $\eta_{\rm dijet}$ refer to the average transverse momentum and rapidity of the two jets with the highest transverse momenta in an event. The CMS data are shown in Figure~\ref{fig:RpPb_dijet} along with next-to-leading oder (NLO) perturbative QCD predictions using the EPPS16 nuclear modifications with CT14 free-proton PDFs~\cite{Dulat:2015mca} and the results after reweighting EPPS16 with the CMS data. What we see is a large reduction in the EPPS16 uncertainties after the reweighting throughout the measured range, showing that the data can put stringent constraints on the nuclear PDFs. Moreover, there is a visible downward pull in the forward region, hinting of a preference for a somewhat stronger shadowing than that given by the EPPS16 central set.

The same conclusions can be drawn from Figure~\ref{fig:EPPS16_rw_dijet}, where the effect on the EPPS16 nuclear modification for gluons in a lead nucleus is shown. We find the data to give additional constraints throughout the probed range. The best-constrained mid-$x$ region shows a strong reduction in the uncertainties, providing clear evidence for gluon antishadowing. Similarly, we find preference for the onset of shadowing at small $x$, but the data are able to constrain the nPDFs only in the region $x \gtrsim 0.002$, leaving the very-small-$x$ gluons still unconstrained (the small-$x$ behaviour of the reweighted uncertainty band in Figure~\ref{fig:EPPS16_rw_dijet} being fixed by the EPPS16 para\-metri\-zation). Moreover, it should be noted that even with the enhanced shadowing compared to the EPPS16 central set, the most-forward data points seen in Figure~\ref{fig:RpPb_dijet} are systematically below the reweighted uncertainty band. Therefore, studies with additional constraints (such as the $D^0$ production discussed below) and with more flexible parametrizations are needed to extract the small-$x$ behaviour of the gluon distribution. We note that there remains also an issue in reproducing the individual normalized p+p and p+Pb spectra, a problem which we have found to be associated with the free-proton PDFs~\cite{Eskola:2019dui}.

\vspace{0.25cm}
\section{Impact of LHCb $D^0$ production data within SACOT-$m_{\rm T}$ GM-VFNS}
\label{sec:d0}

\begin{figure}
  \centering
  \includegraphics[width=0.43\textwidth]{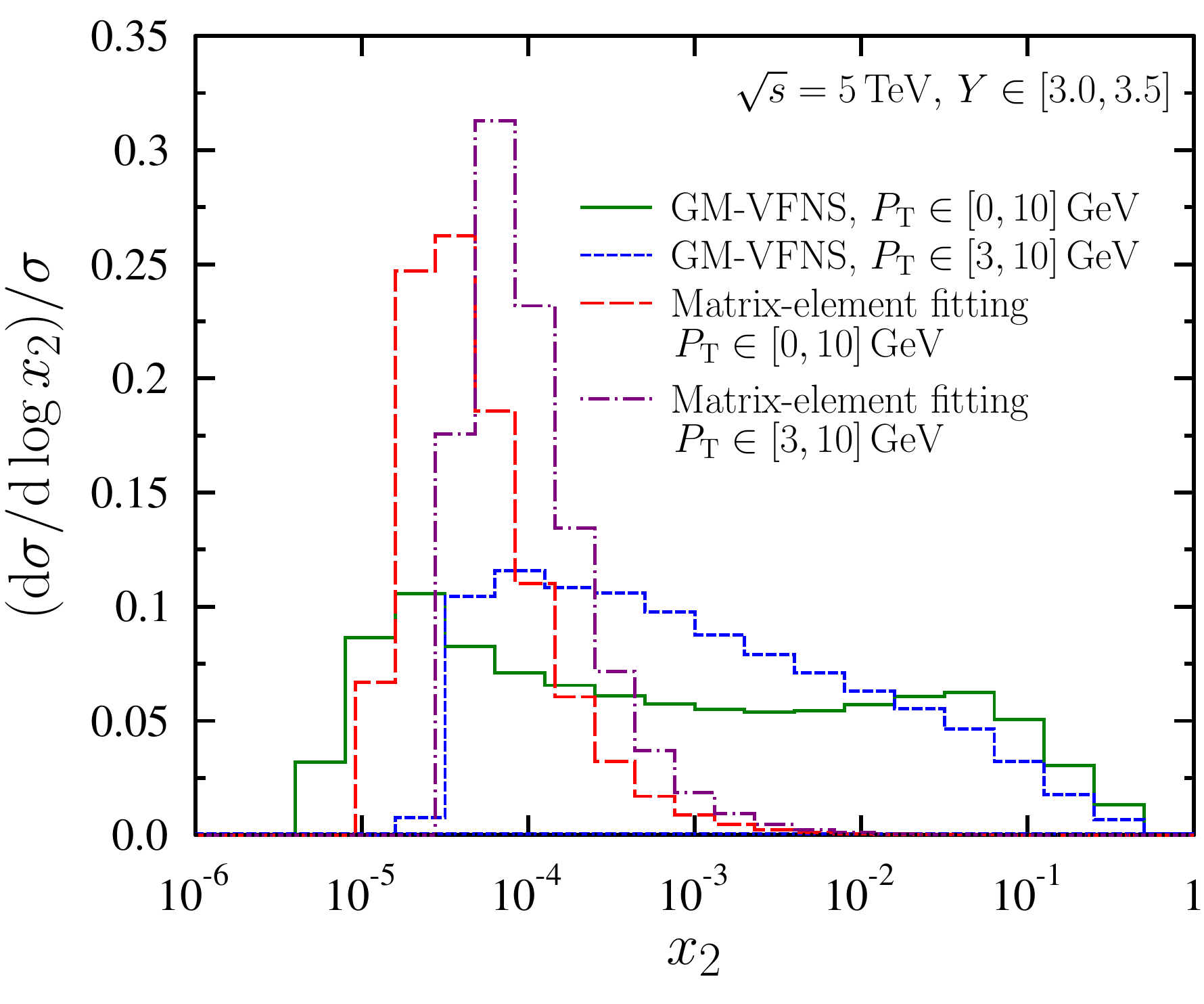}
  \quad
  \includegraphics[width=0.43\textwidth]{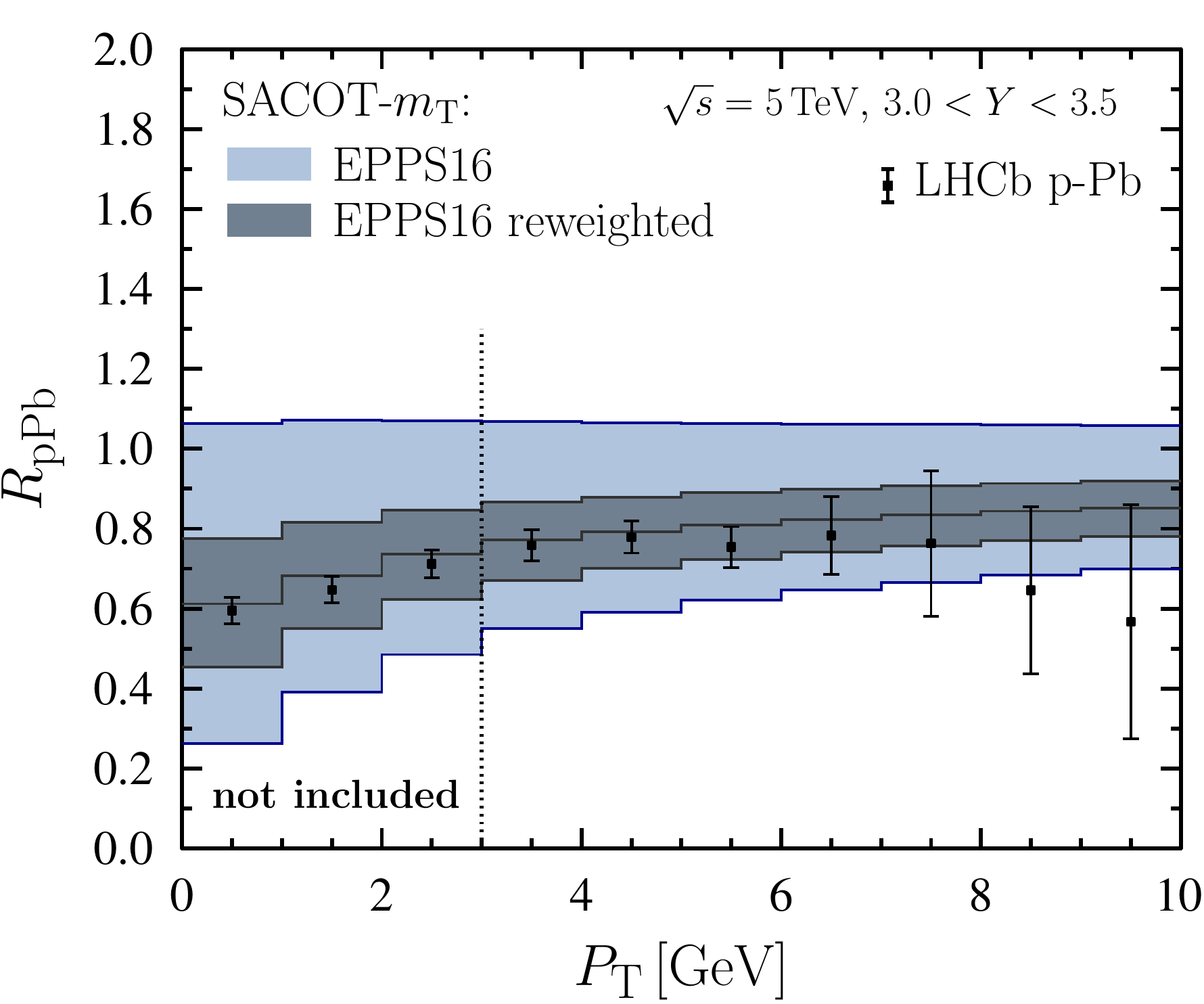}
  \vspace{-0.25cm}
  \caption{\emph{Left:}
  The contributing $x$ regions for $D^0$-production in the SACOT-$m_{\rm T}$ GM-VFNS and a matrix-element-fitting approach in a forward rapidity bin.
  \emph{Right:}
  An example forward rapidity bin of the LHCb $D^0$ nuclear modification ratio and a comparison with the SACOT-$m_{\rm T}$ GM-VFNS calculations using the EPPS16 nPDFs before and after reweighting with the data.
  Figures from Ref.~\cite{Eskola:2019bgf}.}
  \label{fig:RpPb_d0}
\end{figure}

The LHCb measurement of $D^0$ production in p+Pb~\cite{Aaij:2017gcy} includes data at forward and backward rapidities and low transverse momenta, making them a unique constraint for nPDF global analyses. Earlier nPDF comparisons with these data, including those in Refs.~\cite{Aaij:2017gcy,Kusina:2017gkz}, have been based on a matrix-element fitting method~\cite{Lansberg:2016deg}, using $2 \to 2$ kinematics and assuming a simple $|{\cal M}|^2 \propto x_1 x_2$ functional dependence for the matrix elements. Figure~\ref{fig:RpPb_d0} (left) illustrates an important difference in this approach compared to the GM-VFNS calculations presented here. While the $x$ distributions in the SACOT-$m_{\rm T}$ GM-VFN scheme have a long high-$x$ tail, the matrix-element fitting gives very narrow distributions. In the GM-VFNS approach the high-$x$ tail arises mainly from proper resummation of all collinear splittings~\cite{Eskola:2019bgf,Helenius:2018uul}, whereas the matrix-element fitting does not take into account any radiative QCD corrections.

\begin{floatingfigure}{0.65\textwidth}
  \centering
  \includegraphics[width=0.65\textwidth]{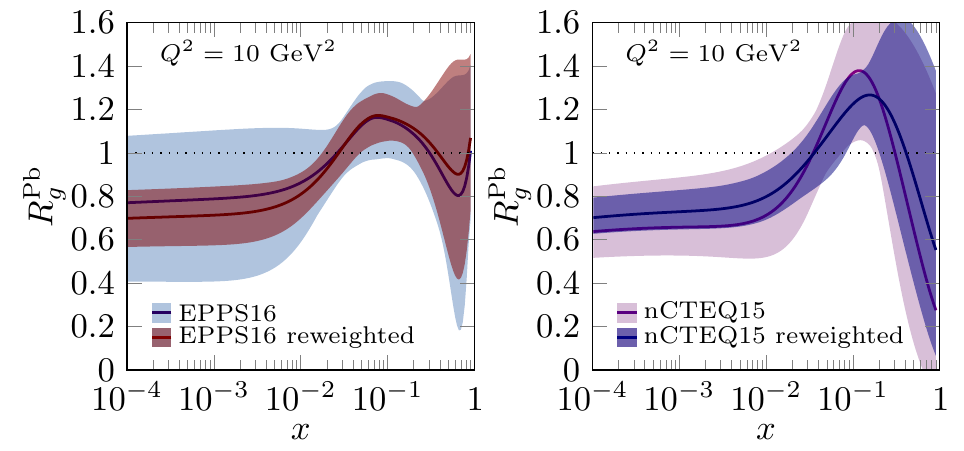}
  \vspace{-0.6cm}
  \caption{The impact of reweighting with the LHCb $D^0$ nuclear modification ratio on the EPPS16 (\emph{left}) and nCTEQ15 (\emph{right}) gluon PDF modifications. Figure from Ref.~\cite{Eskola:2019bgf}.}
  \label{fig:nPDFs_rw_d0}
\end{floatingfigure}

Figure~\ref{fig:RpPb_d0} (right) compares the measured nuclear modification ratio, defined as
\begin{equation}
  R_{\mathrm{pPb}}
  = \frac{\frac{1}{208}\mathrm{d}\sigma^{\mathrm{p+Pb}\rightarrow D^0 + X} / \mathrm{d}P_{\mathrm{T}} \mathrm{d}Y}{\mathrm{d}\sigma^{\mathrm{p+p}\rightarrow D^0 + X} / \mathrm{d}P_{\mathrm{T}} \mathrm{d}Y},
\end{equation}
where $P_{\mathrm{T}}$ and $Y$ are the transverse momentum and rapidity of the produced $D^0$ meson, in a forward rapidity bin to a NLO calculation in the SACOT-$m_{\rm T}$~\cite{Helenius:2018uul} GM-VFNS with the EPPS16 nuclear and the CT14 free-proton PDFs. Also the effect of reweighting is shown. In this case, we have shown that the theoretical uncertainties from scale variation and fragmentation variable definition efficiently cancel in this ratio. However, at small $P_{\rm T}$ these begin to grow and are harder to reliably quantify, for which reason we have excluded the $P_{\rm T} < 3 {\rm\ GeV}$ from the fit. Even with the $P_{\rm T}$ cut in place and taking into account the high-$x$ tail, these data can probe nPDFs down to $x \sim 10^{-5}$, as can be seen also from  Figure~\ref{fig:RpPb_d0} (left).

The data are remarkably precise, and the EPPS16 uncertainties in Figure~\ref{fig:RpPb_d0} (right) get strongly reduced in the reweighting. This directly translates to strong additional constraints on the gluon nuclear modification, as seen in Figure~\ref{fig:nPDFs_rw_d0} (left). Similarly to the dijet case, there is a preference for somewhat stronger shadowing than in the EPPS16 central set, but here additional constraints are more stringent and extend to far smaller $x$, giving a direct evidence of nuclear gluon shadowing. The data also favour mid-$x$ antishadowing, but the reduction in the uncertainty is not as significant as with the dijets. We have repeated the analysis using the nCTEQ15 nPDFs~\cite{Kovarik:2015cma}, with the results shown in Figure~\ref{fig:nPDFs_rw_d0} (right). The reweighting yields similarly good results as with EPPS16. In this case, the data favour somewhat less pronounced shadowing and antishadowing than originally present in the nCTEQ15 central set, and the end result is very similar to that obtained in reweighting EPPS16. The data are thus compatible with both parametrization choices and can help to bring the gluon modifications in the two analyses to a good mutual agreement.

Intriguingly, also the low-$P_{\rm T}$ data points which were excluded from the reweighting are well reproduced with the reweighted nPDFs, as can be seen in the case of EPPS16 from Figure~\ref{fig:RpPb_d0} (right). This corroborates the validity of the used GM-VFN scheme in the whole data range, even down to zero transverse momenta, albeit with the theoretical uncertainties discussed in Refs.\ \cite{Eskola:2019bgf,Helenius:2018uul}. Also, since the low-$P_{\rm T}$ data are well described in terms of purely DGLAP evolved parton densities, we
find no evidence of nonlinear evolution effects here.

\vspace{-0.25cm}
\section{Conclusions}
\label{sec:concl}

We have discussed here the impact of the CMS dijet~\cite{Sirunyan:2018qel} and LHCb $D^0$~\cite{Aaij:2017gcy} nuclear modification ratio measurements on nPDFs in the light of Hessian PDF reweighting~\cite{Paukkunen:2014zia}. As is evident from comparing Figures~\ref{fig:EPPS16_rw_dijet} and \ref{fig:nPDFs_rw_d0}, the constraints obtainable from these data are driving the nPDFs to the same direction. This lends support to the universality of nuclear PDFs in hard-scattering processes and shows that one should be able to include both data sets simultaneously in a global analysis without any substantial tension. Together, these data sets give compelling evidence of mid-$x$ gluon antishadowing and small-$x$ shadowing, with the $D^0$ data extending the constraints on the latter down to $x \sim 10^{-5}$.

\paragraph*{Acknowledgments}
The Academy of Finland projects 297058 (K.J.E.) and 308301 (I.H. and H.P.) are acknowledged for financial support.
P.P.~is supported by Ministerio de Ciencia e Innovaci\'on of Spain under project FPA2017-83814-P; Unidad de Excelencia Mar\'\i a de Maetzu under project MDM-2016-0692; ERC-2018-ADG-835105 YoctoLHC; and Xunta de Galicia (Conseller\'\i a de Educaci\'on) and FEDER.
We thank the Finnish IT Center for Science (CSC) for the computational resources allocated under the project jyy2580.

%% References


\bibliographystyle{elsarticle-num}
\bibliography{nupha-QM2019.bib}






% \end{thebibliography}

\end{document}